\documentclass{emulateapj}
\usepackage{natbib,color,ulem,amsmath}
\usepackage{subfig}

\def\refnew#1{(\ref{#1})}
\def\be{\begin{equation}}
\def\ee{\end{equation}}

\def\yr{\, \rm yr}
\def\s{\, \rm s}
\def\km{\, \rm km}
\def\cm{\, \rm cm}
\def\m{\, \rm m}

\def\g{\rm g}
\def\vhill{v_{H}}

\defcitealias{GLS}{GLS}
\defcitealias{SS11}{SS11}

\shorttitle{Collisionless Growth of Planetesimals}
\shortauthors{Shannon et al.}

\begin{document} 

\title{\mbox{Conglomeration of kilometre-sized planetesimals}}

\author{Andrew Shannon$^{1,3}$, Yanqin Wu$^{1}$ \& Yoram Lithwick$^{2}$}
\affil{$^1$Department of Astronomy and Astrophysics, University of Toronto, Toronto, ON M5S 3H4, Canada;}
\affil{$^2$Department of Physics and Astronomy, Northwestern University,
Evanston, IL 60208 and
Center for Interdisciplinary Exploration and Research in Astrophysics
(CIERA)}
\affil{$^3$Institute of Astronomy, University of Cambridge, Madingley Road, Cambridge CB3 0HA, UK}

\setcounter{equation}{0}
\setcounter{figure}{0}
\setcounter{table}{0}

\begin{abstract} 
  We study the efficiency of forming large bodies, starting from a sea
  of equal-sized planetesimals.  This is likely one of the earlier
  steps of planet formation and relevant for the formation of the
  asteroid belt, the Kuiper belt and extra-solar debris disks.  Here
  we consider the case that the seed planetesimals do not collide
  frequently enough for dynamical collisional to be important (the
  collisionless limit), using a newly constructed conglomeration code,
  and by carefully comparing numerical results with analytical
  scalings. In the absence of collisional cooling, as large bodies
    grow by accreting small bodies, the velocity dispersion of the
    small bodies ($u$) is increasingly excited.
  Growth passes from the well-known run-away stage (when $u$ is higher
  than the big bodies' hill velocity) to the newly discovered
  trans-hill stage (when $u$ and big bodies both grow, but $u$ remains
  at the big bodies' hill velocity).  We find, concurring with the
  analytical understandings developed in \citet{2014ApJ...780...22L},
  as well as previous numerical studies, that a size spectrum $dn/dR
  \propto R^{-4}$ results, and that the formation efficiency, defined as
  mass fraction in bodies much larger than the initial size, is $\sim
  {\mathrm{ a \,\, few}}\times R_\odot/a$, or $\sim 10^{-3}$ at the
  distance of the Kuiper belt. We argue that this extreme
    inefficiency invalidates the conventional conglomeration model for
    the formation of both our Kuiper belt and extra-solar debris
    disks. New theories, possibly involving direct gravitational
    collapse, or strong collisional cooling of small planetesimals, are
    required.
\end{abstract}

\section{Introduction}

Large solid bodies, ranging in size from less than a kilometer to
thousands of km roam in the outskirts of both the solar and many
extrasolar systems. In the case of the solar system, Pluto, Sedna and
other large Kuiper belt objects are directly detected
\citep{PlutoDiscovery,
  1993Natur.362..730J,2005ApJ...635L..97B,2009Natur.462..895S}; and in
the case of extrasolar ones, these bodies are inferred from the dust
grains that are produced during their mutual collisions. Even a small
amount of dust can intercept enough star light to become detectable as
debris disks \citep{Aumann,2008ARA&A..46..339W}.  Surprisingly,
$\,\,\sim 20\%$ of solar-type stars, even at an age similar to the
Sun, harbor dust disks that are brighter than ours by more than three
orders of magnitude \citep{Meyer}.

How do these large bodies form? Are they an intermediate step in the
formation route for planets? Why are they not incorporated into Earth
or even Neptune-like planets? Why are many exo-debris disks so bright
yet ours so anemic? These are the questions we set out to answer.

The conventional picture for the formation of our own Kuiper belt,
\citep[][hereafter SS11]{Safronov1969,1978Icar...35....1G,WetherillStewart,
  1998AJ....115.2136K,2008ApJS..179..451K,2010Icar..210..507O, SS11}
postulates that it started from $\sim 10$ Earth masses of primordial
rock and ice ($\sim$ solid mass in the Minimum Mass Solar Nebulae)~in kilometre-sized bodies.  These bodies grew due to pairwise
  collisions, but only a fraction $\epsilon \sim 10^{-3}$ of the
  total mass conglomerated into the large ($\sim 100 - 1000$~km)
Kuiper belt bodies we see today.  The rest was somehow removed.  This
extremely low efficiency occurs because, theorists argue, as bodies
like Pluto grow, they stir the orderly orbital motion of the small
bodies to such an extent that the latter can no longer be accreted
quickly into large bodies: Plutos starve themselves.

While the calculations behind this picture appear sound, there are serious issues with applying this theory to realistic systems. In the following, we first discuss why this theory fails to explain the observed extra-solar debris disks, followed by a more detailed discussion on why we believe it fails to explain our own Kuiper belt.

The brightness of Gyrs-old exo-debris disks implies that the total
  mass contained in their large objects must be $\sim 1000$ times that
  in our own system \citep{2011ApJ...739...36S}.  The above
    mentioned low efficiency then demands that the extra-solar debris
    disks be formed out of $\sim 10^4 M_\oplus$ of primordial
  solids.  The total disk mass, if supplemented with the missing
    hydrogen gas, would then reach of order a solar mass. This is very
    surprising and, in our view, invalidates the low-efficiency
    conglomeration model. Moreover, a more top-heavy size distribution
    is required to explain the brightness evolution of these debris
    disks
    \citep[e.g.,][]{2007ApJ...663..365W,2008ApJS..179..451K,Lohne:2008,2013ApJ...768...25G,2014MNRAS.442.3266K},
    than is produced by conglomeration model.

We now turn to discuss the Kuiper belt, with its large sea of
    literature. With a current mass of $0.1 M_\oplus$ in large bodies,
    the Kuiper belt appears to be well explained by the conglomeration of an
    initial disk of $\sim 10 M_\oplus$~in kilometre sized bodies. To be compatible with
    observations, the bulk of these kilometre sized planetesimals had to have been somehow
    be removed, either by grind-down and blown-away by solar radiation
    pressure, or by dynamical interactions with giant planets
    \citep{Malhotra,1999Natur.402..635T}.  It is easy to see that the
    latter removal scenario is not compatible with the existence of the large bodies today. If the dynamical scatterings occurred
    before the formation of large Kuiper belt objects, these objects would have had to from in a much less massive disk, incompatible with
    the low efficiency of $10^{-3}$. In the opposite case,
    where the scatterings occurred after the formation, large KBOs will be removed
    as frequently as the small ones, asthe
    scattering process is mass-insensitive
    \citep{1991Icar...90..271S,2008Icar..196..258L}.  This would imply a much more massive KBO belt (in large bodies) must have originally formed than is observed today, again requiring a high efficiency for their
      formation.  For example, if scattering removed $90\%$ of the
      primordial mass, then the large KBOs would have to have been $10\times$
      more numerous in the past.

What about removal by radiation pressure? While plausible for
      the other trans-Neptunian populations, this is clearly not an
      option for the cold classical Kuiper belt situated at $a \geq
      40$ AU.  These bodies are likely formed in-situ, as evidenced by
      their unique colors and dynamically cold orbits \citep{2011ApJ...739L..60B,2012ApJ...750...43D,2014ApJ...793L...2L}.  And they are likely formed in a disk that is low in
      mass: models of Neptune's migration demands that the massive
      MMSN disk has to end at $\sim 30$ AU
      \citep{1999Natur.402..635T,2004Icar..170..492G}, or else Neptune
      will continue to plow beyond its current position.  Moreover, a
      massive disk of $\sim 10 M_{\oplus}$ of kilometer-sized bodies
      would have disrupted the long period Kuiper belt binaries, which
      are observed in abundance in the cold classical
      belt \citep{2011ApJ...743....1P,2012ApJ...744..139P}.

 So in summary, both the extra-solar debris disks and our own
      Kuiper belt require a formation mechanism that is more efficient
      than the conventional conglomeration model. One such possibility
      may be that these large objects are results of direct
      gravitational collapse within the protoplanetary disk, initiated
      by, e.g., streaming instability
      \citep{YoudinGoodman,Johansen:2007}~or turbulent concentration \citep{2008ApJ...687.1432C,2010Icar..208..505C}
      This possibility is epitomized in a catchy phrase,  `Asteroids are
      born big', by \citet{2009Icar..204..558M}.  Unfortuantely, this
      scenario has yet to provide reliable predictions for both the
      efficiency and the size
      distribution of the objects thus formed. 

 A second approach is to boost the efficiency by cooling the seed
  planetesimals, either by gas damping
  \citep{Rafikov2003b,2010A&A...520A..43O,2012A&A...544A..32L,2012A&A...540A..73W}, or by inelastic collision
  between the seeds \citep[][hereafter GLS]{GLS}.
 In a companion paper, we explore the collisional cooling in
  detail. This is important when planetesimal seeds are so small that
  their frequent mutual collisions are able to keep their velocity
  dispersion low.
  This then guarantees efficient accretion by large bodies, until
  almost all mass has been converted to large bodies.  Such a
  high-efficiency scenario allows the {\it in situ} formation of the
  cold classical Kuiper belt, as well as the high mass in bright
  exo-debris disks.

  But for the current publication, we restrict ourselves to the
  conventional case of no collisional cooling. There are three
  purposes to this paper. One is to document our coagulation code and
  certify its various components against analytical expectations (\S
  \ref{sec:code}). This code contains collisional break-down and
  cooling, but these two effects are unimportant here since we adopt
  large initial sizes for the planetesimals.  Second, we
    vigorously demonstrate that, in the case of collisionless growth,
  the formation efficiency is limited to $\epsilon \sim {\rm a\, few}
  \times \alpha$ where $\alpha = R_\odot/a$, the angle subtended by
  the Solar disc at a distance $a$ (\S \ref{sec:simulation}).  In the
  Kuiper belt region, $\alpha\sim 10^{-4}$.  Hence only $\epsilon \sim
  10^{-3}$ of the initial mass could have been incorporated into large
  bodies that one observes there today (if growth was nearly
  collisionless). Equivalently, to form these bodies the initial mass
  in planetesimals would have had to be $\sim 10 M_\oplus$, similar to
  that expected in a MMSN disk extrapolated to the Kuiper belt
  distance, \citep{Weidenschilling:1977,Hayashi}).  Lastly, we compare
  results from our code against previous work (\S
  \ref{sec:compare}). In particular, \citetalias{SS11} have recently
  proposed a simple explanation for the size distribution of Kuiper
  belt bodies, in the case of collisionless conglomeration. We examine
  their claim critically.

This paper should be considered a numerical companion to the recent
work by \citet{2014ApJ...780...22L} where a conglomeration phase called
`trans-hill growth' is first discussed.  That paper, largely
analytical, arrives at similar conclusions as those reached here.

\section{The Conglomeration Code: Component testing}
\label{sec:code}

We adopt a particle-in-a-box approach \citep{Safronov1969} to study
the interactions of planetesimals and their growth. The range of body
sizes we track runs from $\sim \km$ to $\sim 10^3 \km$.\footnote{This
  is expanded in our next work when we consider grains as small as
  $\sim \micron$.}  Since the total number of particles active in our
simulations can exceed $10^{15}$, and we typically
integrate the system for up to $10^6$~dynamical times, we adopt by
necessity a statistical approach where particles with similar
properties (mass, eccentricity, inclination, semi-major axis) are
grouped into the same bin.  We do not yet have the capability of
coupling N-body integrators with the statistical code, as was done by
\citet{2006AJ....131.2737B,2008ApJS..179..451K,2006PhDT.........3G}. As such, we
cannot accurately capture the evolution at late times, when only a few large bodies
dominate the evolution.
 
\subsection{Particles and Bins}

A total of six parameters are needed to specify an orbit. We choose
the semi-major axis $a$, the eccentricity $e$, the inclination $i$,
the argument of periapse $\omega$, the longitude of the ascending node
$\Omega$, and the mean anomaly $M$. Together with mass, each particle
should be described by seven parameters.  This is computationally
prohibitive and approximations are in order. In the following we
explain the simplifying assumptions we make in this code.  

We adopt a single-value of semi-major axis ($a=45$AU) for all
particles, spread to a width $\Delta a \sim 0.13 a \sim 6$ AU.
Our single zone approach limits us to study systems where the
eccentricities are small ($e < \Delta a/a$).  The particles are
assumed to be distributed uniformly in $\omega$, $\Omega$,~and $M$.
We  set $i=0.5e$, as is approximately true if random 
energy is equipartitioned in each dimension
\citep{1985Icar...64..295H}. So velocity anisotropy, important when
particles are very cold \citep{1992Icar...96..107I,Rafikov2003b}, is
not properly treated here.

These simplifications allow us to describe particles only by their
masses and eccentricities.  Following the approach of
  \citet{WetherillStewart}, particles are first binned by mass.
Mass bins are spaced logarithmically, typically
with  20 bins per mass decade (i.e., a relative width $\Delta m/m \approx 1.12 $).  
 These mass bins may ``float'' around as evolution proceeds,
  allowing
 for more accurate tracking of the mass distribution.  Variations of this
  approach are commonly used in conglomeration codes as they allow for
  reasonably fast computation times
  \citep[e.g.,][]{1998AJ....115.2136K}.  Each mass bin is assigned a
  single eccentricity value, which evolves with time.  Particle radii
are computed by assigned all particles the same bulk density,
$\rho=1.5 \g/\cm^3 $.  

The dispersion velocity of a particle is \citep{Wetherill:1993},
\begin{equation}
v = v_{\rm{kep}} \sqrt{\frac{5}{8}e^2+\frac{1}{2}i^2}\,
\label{vimpactequation}
\end{equation}
where $v_{\rm kep} = \Omega a$ with $\Omega$ being the orbital angular
frequency.

Processes detailed in the following section drive mass and
eccentricity evolution and we use a second-order Runge-Kutta
integrator to advance the simulation in time.  The timestep is set so that no bin has an eccentricity change of more than $\Delta e/e = 0.01$, and the change in the total number of bodies in a bin is not more than $\Delta n/n = 0.01$, although $\Delta n$~is always allowed to be $1$.

 There are a few technical subtleties regarding mass evolution.
 When the mass of a bin strays by more than half a bin spacing from
  its original value, bodies are promoted to the next highest bin,
  while the mass of that bin is reset.  This is done in a way to
  conserve the total mass and the total number of particles.  As the average mass of a promoted body will not exactly match the average mass in the next higher mass bin, this slightly alters the average mass of the next higher mass
  bin.  Furthermore,  
for the largest few bodies, their respective zones of dynamical
  influence do not overlap, and they should not interact with one another.  The size of such bodies is called the ``isolation mass'' (size $R_{\rm iso}$)
\citep{1992Icar...96..107I,Wetherill:1993,1998Icar..131..171K} , defined by
\begin{equation}
 \sum_{R=R_{iso}}^{R_{max}} 3.5R_H + 2ae \geq \Delta a\, .
\end{equation}
 We emulate the approach of \citet{1998AJ....115.2136K},
 by pseudo-isolating the biggest bodies so that they do not accrete
each other.  We also require that bodies larger than $R_{iso}$~to be promoted in integer numbers, to avoid the problem of the runaway growth of a fractional body.

\subsection{ Dynamical Friction \& Viscous Stirring}

Mutual gravitational interactions lead towards equipartition of random
kinetic energies between bodies.  This is called dynamical friction
\citep{1943ApJ....97..255C}.  Additionally, in a keplerian disk,
gravitational interactions between bodies converts their orbital
energy into random kinetic energy.  This is called viscous stirring
\citep{Safronov1969}.

\begin{figure}[ht]
\begin{center}
\includegraphics[width=.35\textwidth, trim = 0 0 0 0, clip, angle=270]{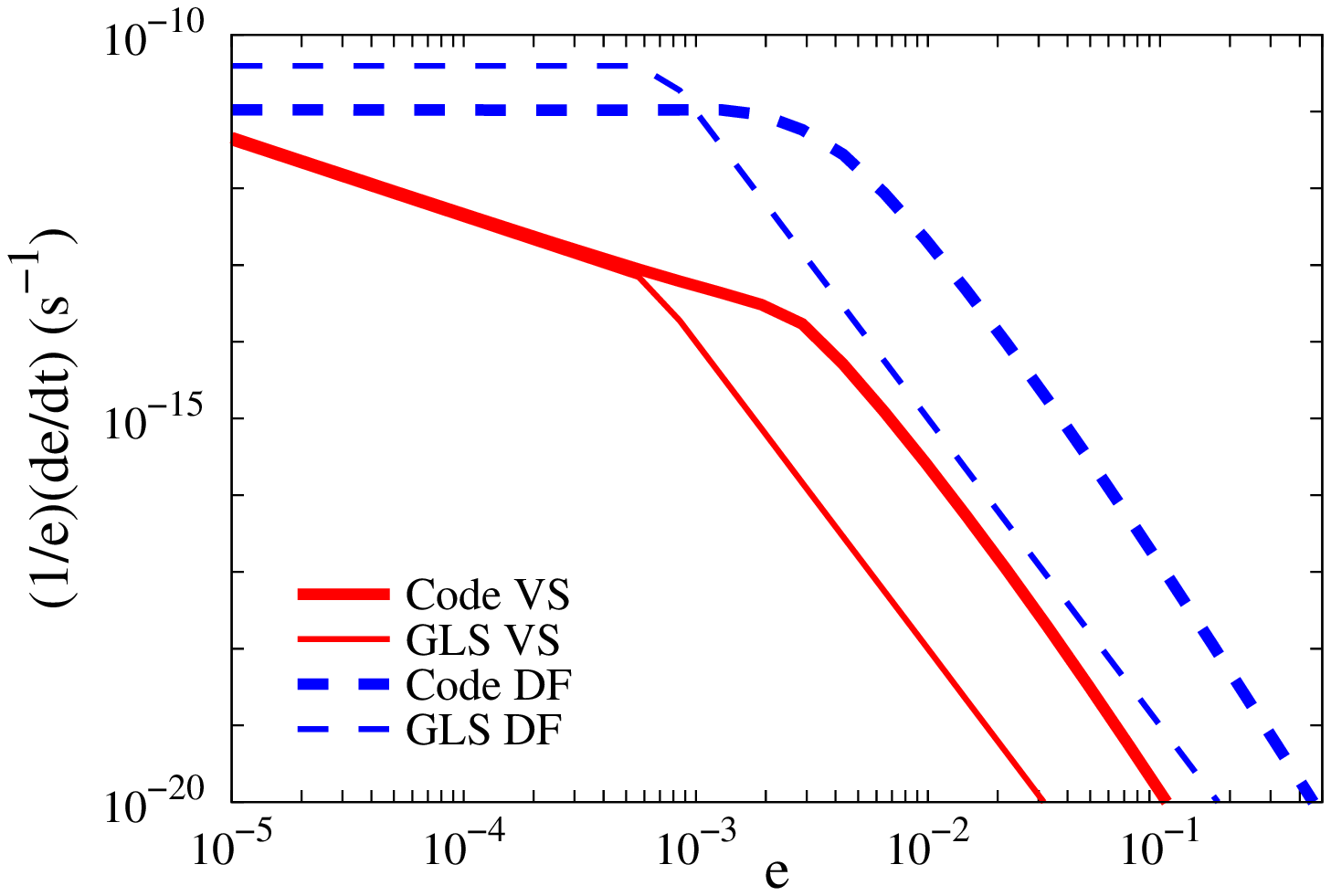}
\caption{ The absolute rates of viscous stirring (VS, red solid lines)
  and dynamical friction (DF, negative in sign and only acting on big
  bodies, blue dashed lines) used in our code (thick lines), as
  compared to the order-of-magnitude estimates (thin lines) of GLS,
  plotted here as functions of eccentricity ($e$).  These rates are
  for a population of $1000\km$ big bodies interacting with a
  population of small bodies, with equal surface densities, $\Sigma =
  \sigma \sim 0.1 \g/\cm^3$, and equal dispersion velocities, $u = v =
  e V_{\rm kep}$, situated at $45$ AU.  The bend around $e \sim
  10^{-3}$ (Hill velocity of the $1000$km bodies) occurs when the
  velocity dispersion transitions from sub-hill to super-hill.  If one
  substitutes $2.5 v_H$ into the GLS expressions whenever $v_H$
  appears, as is suggested by N-body experiments
  \citep{1983PThPh..70...93N,1991Icar...94...98G}, the agreements
  between the two sets of curves will be much closer. }
\label{fig:glscomp}
\end{center}
\end{figure}

To model dynamical friction and viscous stirring, we adopt the
prescription of \cite{Ohtsuki:2002}, who provide semi-analytic formulae
for the rates at which eccentricity and inclination evolve through
gravitational scatterings, calibrated by N-body
simulations.  
But when the mutual velocity of two bodies are below their mutual Hill
velocity (sub-hill), we adopt instead the viscous heating prescription
in eq. (13) of \citet{2007AJ....133.2389C}, calibrated by the
numerical simulations of \citet{2006AJ....132.1316C}.

GLS give explicit formula for the rate of viscous stirring and
dynamical friction, depending on whether the velocities are sub- or
super-hill (also called shear- or dispersion-dominated). They consider
two groups of particles, smaller bodies with size $s$, surface density
$\sigma$ and velocity dispersion $u$, and large bodies with radius
$R$, surface density $\Sigma$ and velocity dispersion $v$.  The two
most relevant expressions are viscous stirring of small bodies by the
big ones, and dynamical friction of the big bodies by the small
bodies. Assuming $u > v$, these are
\begin{eqnarray}
\left.\frac{1}{u}\frac{du}{dt}\right|_{vs} &\sim \frac{\Sigma \Omega}{\rho R}\alpha^{-2}
\begin{cases}
\left(\frac{v_{{H}}}{u}\right)^4 \quad \quad &u > v_{{H}}, \\
\left(\frac{v_{{H}}}{u}\right) \quad \quad &u < \vhill
\label{eq:glsvs}
\end{cases} \\
\left.\frac{1}{v}\frac{dv}{dt}\right|_{df} &\sim -\frac{\sigma \Omega}{\rho R}\alpha^{-2}
\begin{cases}
\left(\frac{v_{{H}}}{u}\right)^4 \quad \quad &u > v_{{H}}, \\
1\quad \quad \quad &u < \vhill
\label{eq:glsdf}
\end{cases}
\end{eqnarray}
Viscous stirring of large bodies by themselves can be found by
substituting $v$~for $u$~in equation \ref{eq:glsvs}. As
Fig. \ref{fig:glscomp} shows, our adopted prescriptions agree with the
above estimates well, when $v_H$ in the above expressions are
substituted by $2.5 v_H$. Here, $v_{H}$ is the Hill velocity of large
bodies,
\begin{equation}
  v_{H} = v_{\rm kep} \left(\frac{M_{\rm big}}{3M_\odot}\right)^{1/3} \approx
3^{-1/3} \alpha^{-1} \, \Omega R\, .
\label{eq:vhill}
\end{equation}
The big body mass $M_{\rm big} = 4\pi/3 \rho R^3$ with $\rho$ being
the bulk density. We set $\rho$ to be the mean density of the Sun, and
use $\alpha$ to denote the angular size of the Sun, viewed by the
bodies, $\alpha = R_\odot/a$. For the Kuiper belt, $\alpha\sim
10^{-4}$.

\subsection{Collisions: cooling and cascade}
\label{chanceofcollision}

Our conglomeration code includes the collisional process. Although we
do not include collisional disruption, and collisional damping is unimportant
in this collisionless study, it is essential for our follow-up
studies of collisional growth. We document both here for completeness.

In the particle-in-box method, the frequency of collision for a
particle is $f_c = n\pi b^2 v$.  Here, $n$~is the number density of
other particles, $\pi b^2$~ the cross section for collision, and $v$~
the relative velocity. 
The relative velocity between two
  particles is simply set to be $v = \sqrt(v_1^2 + v_2^2)$ where $v_i$
  is as defined in equation \refnew{vimpactequation}.

  Under our assumption of evenly distributed orbital angles, particles
  with eccentricity $e$ occupy a torus in space that has a volume
  \citep[see][for a detailed derivation]{Krivov},
\begin{equation}
  V (e) = \frac{4\pi}{3} a^3\left[\left(1+e\right)^3 
    - \left(1-e\right)^3\right]\sin{i}\, .
\end{equation}
Their number density is then obtained by $n=N/V$ where $N$ is the
particle number. Between two groups of particles with different $e$'s
($e_1, e_2$, but the same $a$), their overlapping volume is determined
by the group with the smaller $e$ (let it be $e_1$), and the collision
frequency is reduced by a ratio of $V(e_1)/V(e_2)$ to take account of
the reduced residence time group $2$ particles spend inside $V(e_1)$.

We adopt a three-piece form \citep[][GLS]{1991Icar...94...98G} for the
cross-section for collision between particles of size
$s_1$ and $s_2$, 
\begin{equation}
\pi b^2 = 
\begin{cases}
\pi\left(s_1 + s_2\right)^2 \left(1+\frac{v_{\rm{esc}}^2}{v^2}\right) &\vhill < v, \\
\pi \left(s_1 + s_2\right)^2 \left(\sqrt{6}\alpha^{-\frac{1}{2}} \frac{\vhill}{v}\right) & \alpha^{\frac{1}{2}}\vhill < v < \vhill , \\
\pi \left(s_1 + s_2\right)^2 \left(\sqrt{6}\alpha^{-\frac{3}{2}}\right) &v < \alpha^{\frac{1}{2}}v_{{H}}\, .
\label{eq:crosssection}
\end{cases}
\end{equation}
We follow GLS and name them {\it super-hill, sub-hill, and super-thin}
cases. The last case occurs when the velocity dispersion is
so small the particles can be regarded as infinitely thin in the
vertical direction.
These expressions apply when the velocity dispersion is isotropic ($i
\sim e$).

The outcome of collisions is modelled as changes in mass and
eccentricity.  We calculate the eccentricity evolution rate for
collisions with the GLS order-of-magnitude estimate for this rate,
\begin{equation} \frac{1}{u}\frac{du}{dt} = -
    k\frac{\sigma \Omega}{\rho s} = - k\nu \, ,
\label{eq:colcool}
\end{equation}
where $\nu$ is the collisional frequency.  Numerically, we obtain a
  dimensionless coefficient $k \approx 0.2$, using Monte-Carlo
simulations of equal mass bodies on orbits with the same ($a$, $e$)
and a coefficient of restitution of zero (very inelastic
  collision).

In terms of mass evolution, collisions can lead to catastrophic
destruction, (inelastic) rebound, or conglomeration. Catastrophic
collision is defined where the primary body loses $\geq 50\%$ of its
mass. We specify this to happen if the specific kinetic energy in the
impact, $\frac{1}{2} \frac{m_1 m_2}{(m_1+m_2)^2} v^2$, exceeds the
disruption threshold \citep{2009ApJ...691L.133S},
\begin{equation}
Q^* \approx 500 \left(s_1^3 + s_2^3 \right)^{-1/9} v^{0.8}
+ 10^{-4} \left(s_1^3 + s_2^3 \right)^{0.4}v^{0.8}\, ,
\label{eq:stewartqstar}
\end{equation}
with all numbers in cgs units. This is the scaling for weak
aggregates, and may be appropriate for Kuiper belt bodies.
For km-sized bodies, $Q^* \sim 150
\left(v/1~\rm{cm}~\rm{s}^{-1}\right)^{0.8}~\rm{erg}~\rm{g}^{-1}$. This
corresponds to a disruption velocity of $v \sim 100\cm/\s$, or $e
\approx 10^{-3}$ at $40$ AU.

\begin{figure}[]
\begin{center}
  \includegraphics[width=.45\textwidth, trim = 0 0 0 0, clip, angle=0]{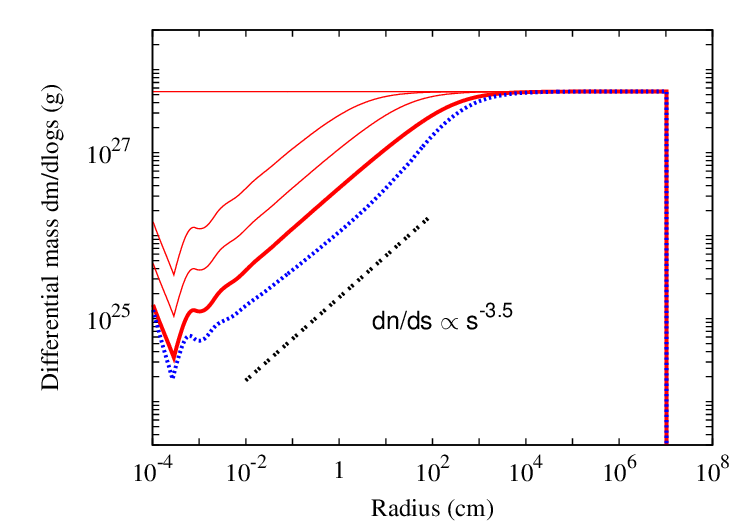}
  \caption{ Evolution of the differential mass distribution during a
    collisional cascade, plotted at $t=0, 10^4, 10^5, 10^6$ yrs (red
    curves).  Material strength is taken to be constant. As a result,
    one expects that bodies in collisional equilibrium satisfy $dn/ds
    \propto s^{-3.5}$ (dotted line) In this simulation, initial
    surface density $\sigma = 0.13 \g/\cm^2$, and mass of the
    catastrophically disrupted bodies are distributed to smaller sizes
    as $dn/ds \propto s^{-2}$.  The final size spectrum is insensitive
    to this debris redistribution: the blue dashed curve shows the
    size distribution at $10^6$ yrs if the redistribution instead
    follows $dn/ds \propto s^{-4}$.  We assume all particles smaller
    than $1 \mu$m are instantly blown away by radiation pressure,
    which leads to the wavy pattern near the cut-off size, as
    described in \cite{2003A&A...408..775T}. }
\label{fig:dohnanyi}
\end{center}
\end{figure}

When bodies of size $s_1$ are catastrophically disrupted, we
re-distribute their masses to
smaller size bins with a number distribution that is power-law in
size, $dn/ds \propto s^{q'}$ for $s < s_1$.
We typically choose $q'=-3.5$. This distribution results in most of the
mass ending up in the largest fragments. And Fig. \ref{fig:dohnanyi}
shows that results do not vary when we vary the value of $q'$ from $-2$
to $-4$. Furthermore, to imitate the effect of radiative pressure, we
remove all mass in size bins below our minimum size (at one micron).
In the future, it may be useful to implement the dependence of the
largest fragment mass on the impact energy
\citep{2009ApJ...691L.133S}.

To verify the collisional mass evolution, we test our code against the
standard case of collisional equilibrium for which an analytical
solution is known. For a material strength $Q^* \propto s^{p}$, the
collisional cascade will carry a constant mass flux downward in
particle size and build up an equilibrium size distribution of
\citep{Dohnanyi:1969,1997Icar..130..140D},
\begin{equation}
\frac{dn}{ds} \propto s^{-q}
\label{eq:coleq}
\end{equation}
with $q = (21+p)/(6+p)$.  The well-known Dohnanyi law is the special
case with $p=0$ and hence $q = 3.5$.  In the numerical experiment
shown in Fig. \ref{fig:dohnanyi}, we adopt a constant strength of $Q^*
= 10^8~\rm{ergs/g}$, and initialize the particle size distribution
with $dn/ds \propto s^{-4}$.  Irrespective of the power-law we
  choose for the debris redistribution, we find that small bodies in
  the system settle into the expected $q=3.5$ form and larger bodies
  gradually enter into collisional equilibrium as time goes on.

If the collision energy is too low to cause destruction, we separate
the outcomes into two further categories: rebound or conglomeration.
In our code, conglomeration occurs when the relative velocity falls
below ten times the mutual escape velocity. This disfavors
conglomeration of small bodies: kilometer-sized bodies only accrete
each other when $e \lesssim 10^{-3}$ at Kuiper belt distances.

When this is not satisfied, the bodies rebound, with an
  eccentricity damping as described by Eq. \refnew{eq:coleq}.  In the
case of unequal masses, the cooling rate is reduced by
$\left(m/M\right) \leq 1$.  We do not model either cratering collision
or sticking by chemical forces.  Note that mass loss due to cratering
collisions may be the dominant source for the destruction of large
bodies \citep{2010Icar..206..735K,2012ApJ...749...14G}, and it may be
necessary to consider them in realistic models of collisional
disruption.  Fragmentation can also reduce the effectiveness of planet
formation,  especially if the fragmenting material can be lost by
other means
\citep{2003Icar..166...46I,2004AJ....127..513K} although this primarily concerns growth at larger sizes than we consider here.
In the following, we discuss in more detail the accretion process.

\subsection{Collisions: accretion}
\label{sec:singlebodygrowth}

The collisional cross sections in eq. \refnew{eq:crosssection} lead to
the following rates of accretional growth (GLS), for bodies with size
$R$ and velocity dispersion $u$, accreting smaller bodies of surface
density $\sigma$ and velocity dispersion $u$ (with $u > v$)
\begin{equation}
\frac{1}{R}\frac{dR}{dt} \sim \frac{\sigma \Omega}{\rho R}\alpha^{-1}
\begin{cases}
\left(\frac{\vhill}{u}\right)^2\, , & \vhill < u < v_{\rm esc}, \\
\left(\frac{\vhill}{u}\right)\, , &\alpha^{\frac{1}{2}}\vhill < u < \vhill , \\
\alpha^{-\frac{1}{2}},\,\,\, &u < \alpha^{\frac{1}{2}}\vhill \, .
\label{eq:allgrow}
\end{cases}
\end{equation}
We confirm that our numerical algorithm does reproduce the above
relations.  There is a subtlety here. These expressions assume an
isotropic velocity dispersion, $i \approx e$. However, when stirring
is in the sub-hill regime ($u < v_H$, eq. \ref{eq:glsvs}),
inclinations of bodies are excited at a lower rate than their
eccentricities, while both quantities are damped by dynamical friction
at comparable rates. We then expect these bodies to satisfy $i \ll e$
and the accretion rate should proceed as the super-thin case (GLS), in
which case the accretion rate is $\sigma\Omega/(\rho R)\cdot
\alpha^{-3/2}$ for all $u<v_H$.  In practice, however, stirring is
contributed by bodies of various sizes, not just the largest
sizes. Short of tracking the inclination evolution, it is difficult to
ascertain the accretion geometry.  So in this work, we simply assume
an isotropic velocity dispersion even though it carries mistakes that
may reduce accretion rate by up to $\alpha^{-1/2}
(u/\vhill)$.  This issue matters
  only when dealing with accertion among big bodies, which,
  fortunately, is not significant in the current study.

\begin{figure*}
  \centering
  \subfloat{\includegraphics[width=0.32\textwidth]{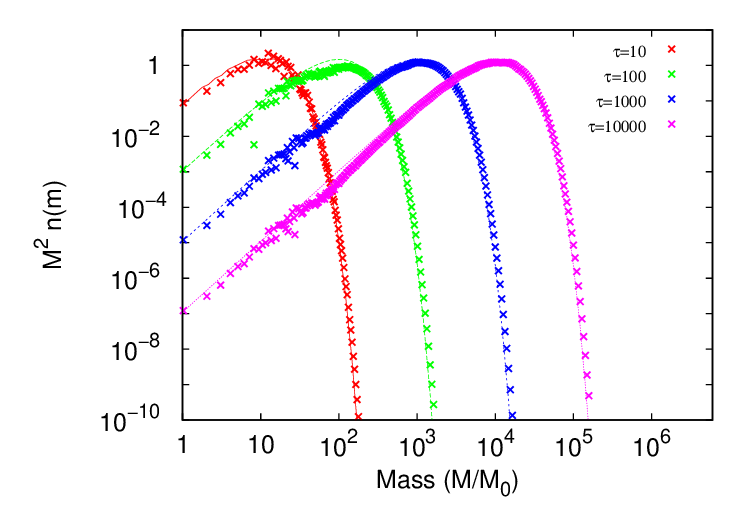}}
  \subfloat{\includegraphics[width=0.32\textwidth]{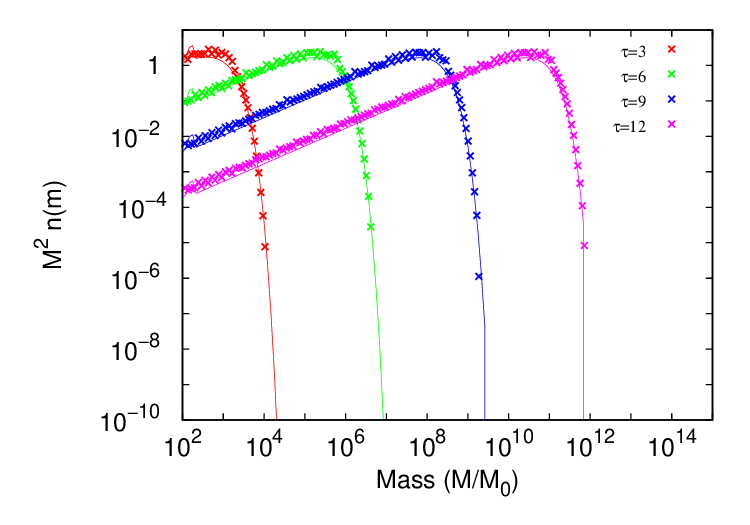}}
  \subfloat{\includegraphics[width=0.32\textwidth]{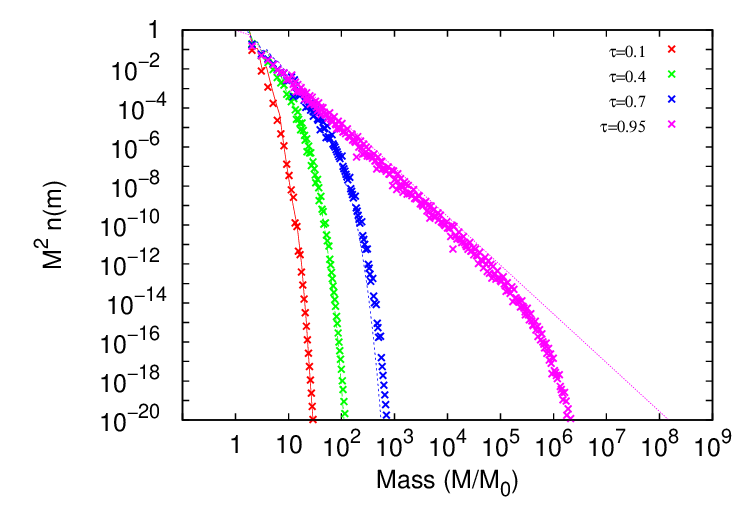}}
  \caption{ Comparison of our code's evolved size-number
      distribution to the analytic solutions for the cases where the
      chance of two bodies colliding is a constant (left),
      proportional to the sum of their masses (centre), and
      proportional to the product of their masses (right).  Here
      $\tau$~is the number of collision times of the smallest bodies.
      Our code reproduces the analytical solutions (thin solid lines) accurately,
      except in the last case where our code reaches the divergence
      at $\tau = 1$ earlier in time than expected (by about $5\%$).}
\label{fig:analtests}
\end{figure*}

We perform tests to verify our accretion
  algorithm. For the following three cases where the accretion cross section between
  any two bodies is, a) a constant, b) scales as the sum of their
  masses, $(m_1+m_2)$, c) scales as the product of their masses, $(m_1
  \times m_2)$, there exist
  analytic solutions for the evolution of the size distribution
  \citep{1990Icar...88..336W}. We integrate these evolutions using our
  code and compare the numerical outcome against the analytical
  results (Fig. \ref{fig:analtests}.  Our code reproduces the analytic
  solutions well, except for case c), where the cross section rises 
  steeply with mass, our numerical results reaches the singularity
  about $5\%$ earlier in time than expected.  Other floating mass bin codes hit this singularity later than expected, not earlier \citep[e.g.][]{1998AJ....115.2136K,2008ApJ...684.1291O}, though the precise disagreement depends on the details of implementation \citep{1990Icar...88..336W}.

\section{Conglomeration: Results and Analysis}
\label{sec:simulation}

\begin{figure*}
  \centering
  \subfloat{\includegraphics[width=0.5\textwidth]{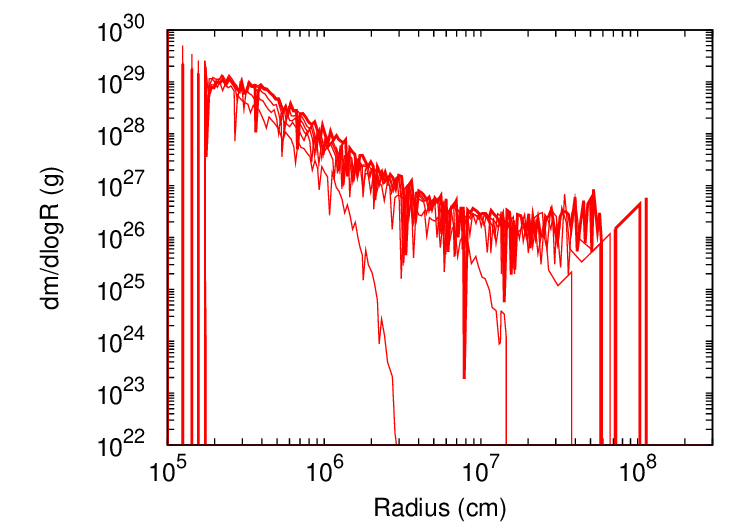}}
  \subfloat{\includegraphics[width=0.49\textwidth]{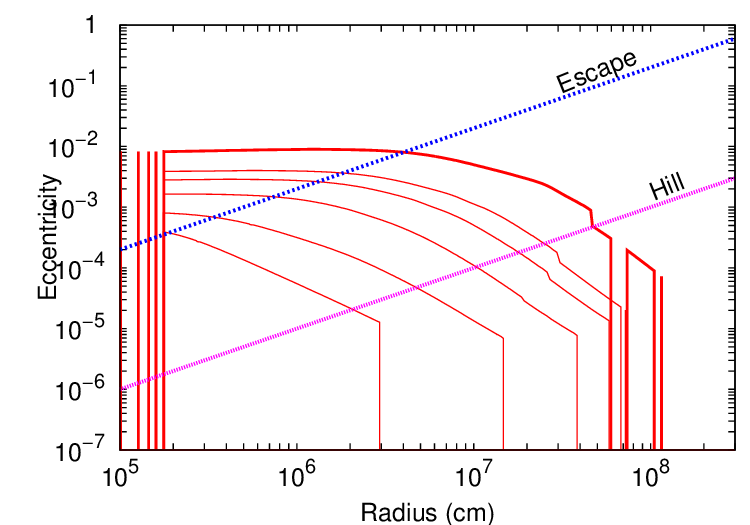}}
  \caption{ Results from our fiducial simulation of collisionless
      conglomeration, starting from a disk of $1\km$ bodies with a
      surface density $\sigma = 0.13 \g/\cm^2$, and an initial
      eccentricity of $e = 10^{-7}$. The left panel shows the
      evolution of differential mass and the right panel eccentricity,
      plotted as functions of body sizes. Data are taken at 0, 20, 40,
      60, 80, 100 (all thin lines) and 200 Myrs (thick line),
      corresponding to $t\nu = 0.0, 0.7, 1.3, 2, 2.7, 3.3, 6.7$.  By
      60 Myrs, the number distribution of large bodies ($100$-$1000
      \km$) can be roughly described by a power-law, $dn/dR \propto
      R^{-q}$ with $q \sim 4$. The efficiency of conglomeration,
      defined here as the mass fraction above $100 \km$, is $\sim
      10^{-3}$. On the right panel, the two dashed lines correspond to
      the surface escape velocity and Hill velocity from bodies of
      that size.}
\label{fig:canoncase}
\end{figure*}

\begin{figure}[h]
\begin{center}
  \includegraphics[width=.45\textwidth, trim = 0 0 0 0, clip, angle=0]{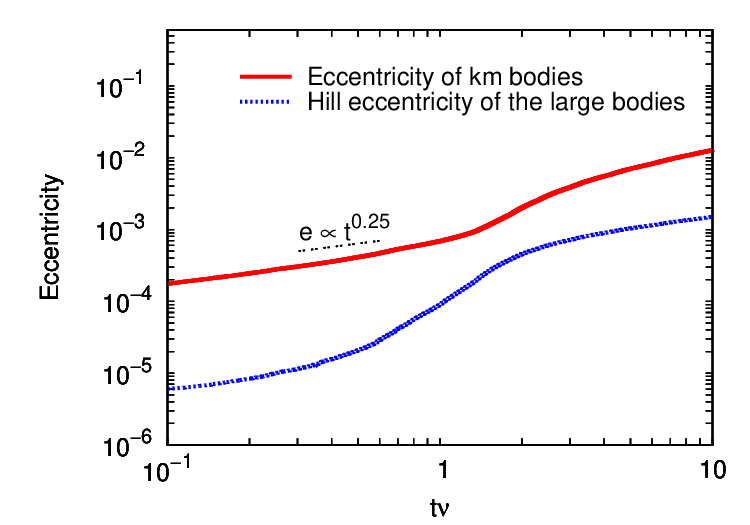}
  \caption{ Evolution of the eccentricity of km planetesimals is
    plotted here against the dimensionless time (solid red curve), for
    the simulation presented in Fig. \ref{fig:canoncase}. The Hill
    eccentricity of the largest bodies at any given time, $e_{\rm
      H}(R_{\rm max})$, is plotted as a blue dashed curve.  In the
    geometric accretion phase ($t\nu \lesssim 1$), the small
    bodies self-stir to produce $e \propto t^{1/4}$. Rapid run-away
    follows once the growth bottle-neck is overcome. In this phase,
    $e$ remains roughly constant as stirring is dominated by the
    inefficient small bodies, while $e_{\rm H} $ rises steeply.  
      After $t\nu \sim 1.3$ the largest bodies have reached $\sim
      100$\km, and the growth of the big bodies is gradually hampered
      by their own stirring, growth proceeds in lockstep with stirring
      with $e \sim 2.5 e_{\rm H}$ (the trans-hill stage).  As growth
      is rapid, the phase is short lived, but critical for setting the
      properties of big bodies.  When the largest bodies have grown to
      $\sim 800-1000$km ($t\nu \sim 3$), oligarchic phase
      commences. Growth effectively stalls and $e$ is again stirred up
      gradually as $e \propto t^{1/4}$.  See text for more details.}
\label{fig:plot172}
\end{center}
\end{figure}

In the remainder of this paper, we present the results of a number of
{\it collisionless} simulations, in which rebounding and destructive
collisions are disabled in the code, but conglomeration still occurs
when bodies collide at a speed less than ten times their mutual escape
speed.  We focus on collisionless simulations in order to compare with
previously published results, in which typically small bodies collide only a few times over the course of the simulation.  In an
upcoming paper, we shall consider the effects of collisional damping
and destruction.

We present detailed results from a fiducial simulation in
Fig. \ref{fig:canoncase}, with Fig. \ref{fig:plot172} providing
diagnostic detail.  For ease of comparison, we initalize our
planetesimal disk similarly to those in previous works
\citep[][SS11]{1998AJ....115.2136K}.  Our disk is made up of $s=1\km$
planetesimals with a surface density of $\sigma = 0.13 \g/\cm^2$,
comparable to the solid value in a MMSN disk.  Spread over a radial
width of $\Delta a/a = 0.13$ at $a = 45$ AU, this corresponds to a
total mass of $10 M_\oplus$. The bodies ($\sim 10^{13}$ of them) are
initially placed on dynamically cold orbits with $e = 10^{-7}$.

The growth of the largest bodies is mainly due to accretion of
kilometer bodies, at all times. So an important quantity to focus on
is eccentricity of the km bodies. The following scalings for the
Kuiper belt region will prove useful: $e_{\rm esc} = v_{\rm
  esc}/v_{\rm kep} \sim 10^{-1} (R/10^8\cm)$, where $v_{\rm esc}$ is
the escape speed from a body of size $R$, and $e_H = \vhill/v_{\rm
  kep} \sim \alpha^{1/2} e_{\rm esc} \sim 10^{-3} (R/10^8 \cm)$.  We
also discuss two related distributions: the number distribution and
the mass distribution. If the number distribution is $dn/dR \propto
R^{-q}$, then the mass distribution is $dm/d\log R \propto R^{4-q}$.

The km bodies have a mean collision frequency 
\begin{equation}
  \nu = \frac{\sigma \Omega}{\rho s}
  \sim 3 \times 10^{-8} \yr^{-1}\, .
\label{eq:nu}
\end{equation}
This is the natural timescale in the problem of interest. So from now on, we express time always in unit of $t\nu$.

\subsection{Three Growth Stages}
\label{subsec:three}

The growth can be naturally divided into a few stages.

{\bf 1. Geometric accretion} Initially, stirring is much faster than
growth until $e\sim e_{\rm esc}$, and hence the initial growth is
dominated by pair-wise conglomeration among equal-sized bodies (size
$s$). Assume collisions occur with geometric cross-section. At time
$t\nu \ll 1$, the average collision probability for each of the
$10^{13}$ km-sized bodies is $(t \nu)$, but the probability of a
single body having experienced $N$ collisions is $(t\nu)^N$.  The size
distribution at this point falls off steeply with $R$.  The largest
body possible at time $t$ is determined by $(t\nu)^N 10^{13} = 1$, or
at a radius of
\begin{equation}
  R_{\rm max} \approx  \left[\frac{-\log(10^{13})}{\log (t\nu)}\right]^{1/3}\, s.
\label{eq:r8first}
\end{equation}
So at $t\nu \sim 0.3$, the largest body has a size of $R \sim\, 3 \km$
(ignoring gravitational focusing; see below).  By this time, km bodies
have stirred each other to $\sim 1/2$ of their escape velocities
(eq. \ref{eq:glsvs}),
or, $e\sim 10^{-4}$, much above $e_H$ of the largest bodies already
formed.

{\bf 2. Rapid Run-away} If small bodies are super-hill, the largest
bodies experience run-away. They pull away from the rest of the pack while
carrying little total mass (left panel of Fig. \ref{fig:canoncase}).
Eq. \refnew{eq:allgrow} indicates that the slowest growth occurs at
the smallest size. So once the geometric accretion produces bodies
with $R \sim 3\times s$, the bottle-neck is overcome and very large bodies
can be conglomerated within a comparable amount of time.

During this stage of rapid runaway, the small bodies dominate the
stirring, and their eccentricities increase very gradually
(see Figure \ref{fig:plot172}).  The big bodies grow faster than they
can stir, and their size spectrum  is very steep, $q \gg 4$.
  With time, $u/\vhill$ decreases towards unity and we enter the
  following newly-discovered phase of growth \citep{2014ApJ...780...22L}.

3. {\bf Trans-hill growth (or slow run-away)} Rapid run-away ends
once the big bodies dominate the stirring.  Thereafter, the big
bodies' stirring interferes with their feeding, and the big bodies
grow at the same rate as the small bodies' eccentricities are
stirred. 
\begin{equation}
  \frac{\Sigma}{\sigma}\alpha^{-1} 
\left(\frac{\vhill}{u}\right)^2 \approx 1\, .
\label{eq:dominantstirrer0}
\end{equation}
 Big body stirring exceeds that of small bodies when 
\begin{equation}
\frac{\Sigma}{\sigma} \geq \left({s\over R}\right)^3\, .
\label{eq:dominantstirrer}
\end{equation}
For our simulation, this is satisfied when $R_{\rm max} \sim 25\km$ and $\Sigma/\sigma \sim 10^{-4}$, at a time of $t\nu \sim 0.7$.

This transition point is affected by the value of $u$, if the
  initial velocity is such that $u > v_{\rm esc}$. If the initial
excitation is $e = 10^{-3}$, for instance, the transition would occur
at $R \sim 10^3 \km$ with again $\Sigma/\sigma = 10^{-4}$.  Lastly,
this transition point can be affected by the initial size spectrum of
big bodies.

Let us define a class of trans-hill bodies ($R_{\rm trans}(t)$) which
always satisfy $\vhill \approx u (t)$. These bodies are $\sim 100\km$
when the rapid run-away transitions to slow run-away. Following
\citet{2014ApJ...780...22L}, we argue that these trans-hill bodies remain
effectively the main stirrers in the system, $R_{\rm stir} \approx
R_{\rm trans}$, or equivalently, $u \sim \vhill (R_{\rm stir})$, in
subsequent growth. Hence the name `trans-hill' growth.

To demonstrate this must be true \citep[for a more detailed
explanation, see][]{2014ApJ...780...22L}, we consider the growth of bodies
above and below $R_{\rm trans}(t)$. Growth of bodies with $R < R_{\rm
  trans}$ is in the super-hill (runaway) regime.
  Their growth rate (eq. \ref{eq:allgrow}) can be written as
\begin{equation}
\frac{1}{R}\frac{dR}{d(t\nu)} \approx \frac{s}{R_{\rm trans}}\alpha^{-1} 
\times
\frac{R}{R_{\rm trans}}
\label{eq:smallgrow}
\end{equation}
 Since this decreases with decreasing $R$, the size spectrum for $R<R_{\rm trans}$
is largely frozen.

Growth of bodies with $R > R_{\rm trans}$ is in the sub-Hill (orderly)
regime.  The shape of their size spectrum remains invariant with time.
Since the initial size spectrum falls steeply beyond $R>R_{\rm
  trans}$, stirring is initially dominated by bodies of size $R_{\rm
  trans}$. And this will remain so throughout the trans-hill growth.

As a result of these two considerations, $u \approx v_{\rm
  H}(R_{\rm stir})$ during the trans-hill growth. Equation
(\ref{eq:dominantstirrer}) continues to hold, so the size spectrum
extends to larger and larger sizes with $\Sigma \sim {\rm const}$, or
$q \approx 4$. This is the value we find in our
simulations, and similar results were found by \citet{1998AJ....115.2136K},
\citet{2010Icar..210..507O}, and \citetalias{SS11}. An explanation for
the slope of the size spectrum, different from our above one, is given
in \citetalias{SS11}.  We discuss our different interpretations further
in \S \ref{sec:comparison}.

In our simulations, bodies as large as $1000\km$ are formed by $t \nu
\sim 3$. This is consistent with the mass-doubling time for the
stirring bodies one obtains from eq. \refnew{eq:allgrow}, $t\nu \sim
\alpha R/s$.  At this time, large bodies are separated too much to
  affect each other \citep[][]{WetherillStewart}, they are oligarchs
  \citep[see, e.g.][]{1998Icar..131..171K}. The appearance of oligarchs
  effectively terminate the trans-hill stage. However, much of the
  growth has occurred.

\subsection{End of Growth}

The efficiency of formation, defined here as the mass fraction 
in bodies larger than $100\km$, reaches $\sim 10^{-3}$ towards the end
of trans-hill growth (Fig. \ref{fig:canoncase}). The largest bodies
formed are $\sim 1000 \km$, or Pluto-sized. We argue below that the
growth effectively stops when these values are reached. In other
words, collisionless conglomeration produces large bodies but at a
very low efficiency. 

During slow run-away, equation \refnew{eq:dominantstirrer} is
continuously satisfied, yielding $\Sigma/\sigma \sim \alpha$.  So the
largest size bodies that can form in our simulation, insisting on at
least one such body, is $\sim 1000\km$.  And indeed we observe that
evolution practically stalls when $R_{\rm max} \approx 1000\km$. The
formation efficiency, defined as the fractional mass accumulated in
bodies greater than $100\km$, is then
\begin{equation}
\epsilon = \frac{1}{\sigma}\, \int_{\rm 100\km}^{R_{\rm max}} \Sigma \, d\log R \approx 
{\rm a\,\,\, few}\,\,\,\,\alpha \ ,
\label{eq:efficiency}
\end{equation}
and is of order $10^{-3}$ in the Kuiper belt region.  In
\citet{2014ApJ...780...22L}, we argue that trans-hill growth is typically
terminated when a single big body dominates its own feeding zone, i.e., when the
biggest bodies become oligarchs \citep{1998Icar..131..171K}.

Our statistical code loses some validity once the oligarchic
phase commences. A statistical code coupled with N-body dynamics is
needed. However, we can project the subsequent growth fairly robustly
for the Kuiper belt region, because there is little growth past this
point. Neither the big body number nor their spectrum will change much
after this.

After trans-hill growth is terminated, small bodies are stirred to
super-hill velocites, as $u \sim t^{1/4}$ until $u$ reaches $v_{\rm
  esc} (R_{\rm max})$ (Fig. \ref{fig:plot172}). With such a hot
population of small bodies, accretion by the big bodies proceeds with
geometrical cross-section.  Yet geometric accretion is far too
inefficient to form the observed population of Kuiper belt objects
within the Solar system lifetime.\footnote{Pluto, for instance, has a
  geometrical optical depth of $(1000\km/40{\rm AU})^2 \sim
  10^{-14}$.}  So there is no subsequent accretion to raise the
efficiency of big body formation much above $\sim 10^{-3}$.  This low
formation efficiency is a consequence of our neglect of collisions
amongst small bodies.  In a forthcoming paper, we show that when
collisions are included, the formation efficiency can rise to nearly
100\%.
  
Growth is likely to evolve into the collisional regime.  Even if
planetesimals start as km-sized, as they are continuously stirred, they
approach their destruction speed at $e \sim 10^{-3}$
(eq. \ref{eq:stewartqstar}). Their mutual collisions beyond this point
should be destructive and will produce many small debris. When this
occurs, collisional cooling among small planetsimals can no longer be
ignored.  Growth will then be in the collisional limit.

  A minor caveat concerns our assumption of isotropic velocity
  dispersion for all bodies.  For large bodies which stir each other
  in the sub-hill range, the inclination dispersion may fall much
  below that of their eccentricities \citep{Rafikov2003b,GLS}. It is
  likely that accretion among large bodies could proceed then
  at a rate greater than that adopted here. 

\subsection{Simulations with Different Initial Conditions}
\label{subsec:vary}

Our above understanding of the conglomeration process allows us to
explain the following dependences when initial conditions in the
simulation are altered. In all cases, we find a similar size spectrum
to that obtained above.  We delay the discussion on the
$\alpha$-dependence to \S \ref{subsec:SS11}.

\begin{itemize}

\item {\bf initial eccentricity} We compare the growth of bodies with
  initial eccentricities ranging from $10^{-7}$ to $10^{-3}$, with all
  other properties held constant. The escape velocity from km bodies
  is $e\sim 10^{-4}$ at 45 AU. We find that simulations with initial
  velocity dispersion smaller than this quickly converge to our
  standard case; while those with velocity dispersion above this value
  first undergo orderly growth and produce a more significant mass in
  intermediate class bodies. In all case, the final  efficiency
    is $\epsilon \approx 10^{-3}$.

\item {\bf initial surface density} We vary the initial surface
  density of solids from $10^{-3}$ times to $10$ times the MMSN values
  (our standard case being $1$ MMSN), with an
  initial eccentricity of $e = 10^{-7}$.
  We observe that the growth time scales inversely with the surface
  density (Fig. \ref{fig:sigmavaryfigure1}). This is because the
  natural timescale in the problem is the small body collision time
  (eq. \ref{eq:nu}). This result has been reported by
  \citet{1998AJ....115.2136K}. We also find that higher mass disks can
  harbor larger $R_{\rm max}$ -- the final size of the largest body
  scales roughly as $\sigma^{1/3}$, a result of us insisting that the
  largest bin has at least one whole body. Lastly, the efficiency of
  formation only depends logarithmically on the initial $\sigma$.

\begin{figure}[h]
\begin{center}
  \includegraphics[width=.45\textwidth, trim = 0 0 0 0, clip, angle=0]{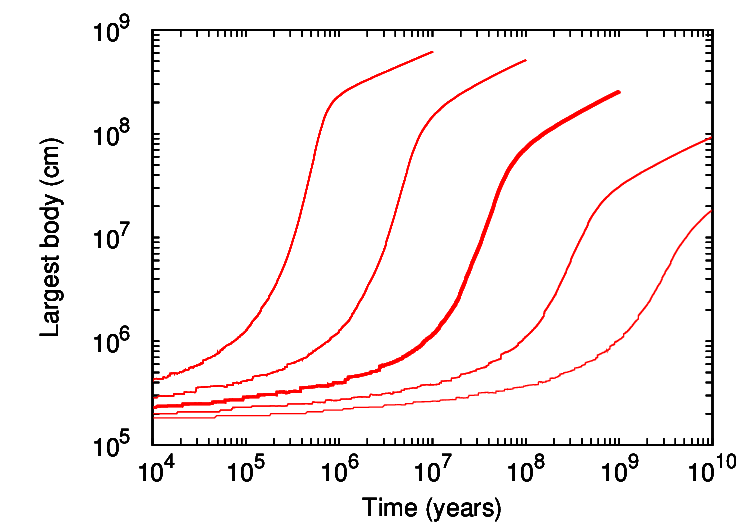}
  \caption{Growth of $R_{\rm max}$ as a functions of time, for disks
    with different surface densities. The total mass spans from
    $10^{-2}$ MMSN (bottom-most curve) to $10^{-1}$, $1$, $10$, 
    and $100$ MMSN (top-most curve), with the thick red curve being our
    fiducial case (Fig. \ref{fig:canoncase}). We find the timescale of
    runaway growth obeys $t\propto \sigma^{-1}$. All growth follow the
    pattern of a rapid run-away, followed by a slow run-away.}
\label{fig:sigmavaryfigure1}
\end{center}
\end{figure}

\begin{figure}[h]
\begin{center}
  \includegraphics[width=.45\textwidth, trim = 0 0 0 0, clip, angle=0]{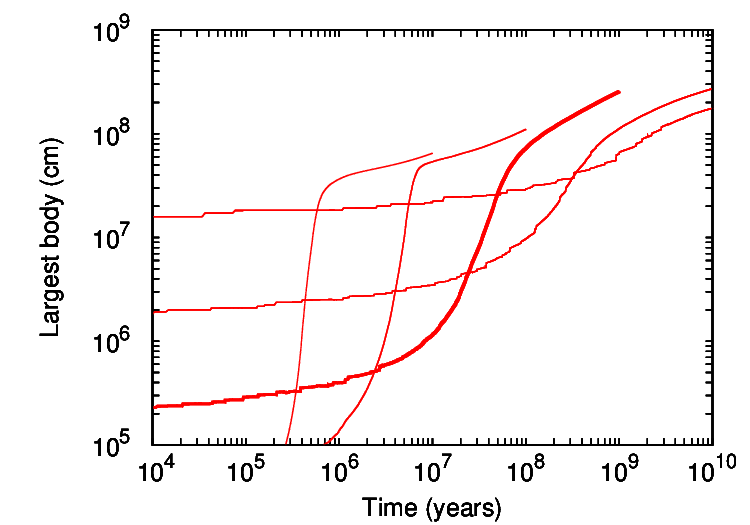}
  \caption{Growth of the largest body in simulations with different
    initial sizes.  Initial sizes range from 10 m (the fasest growing)
    to 100 km (the slowest growing).  Bodies grow slowly at first,
    with the velocity dispersion set by the escape velocity of the
    initial bodies.  Eventually a few bodies run away, when they are
    sufficiently large.  The timescale of runaway increases linearly
    with starting size. However, the largest bodies that formed are
    similar in size.}
\label{fig:plot160}
\end{center}
\end{figure}

\item {\bf initial size} We perform a suite of simulations with
  different starting sizes ranging from $10$~m to $10$~km. Since the mean
  collision time (eq. \refnew{eq:nu}) scales inversely with starting
  size, and the entire evolution takes place within a few collision
  times, we expect that a smaller starting size leads to quicker growth.
  This is observed in Fig. \ref{fig:plot160} where the growth time
  scales roughly as $ \nu^{-1} \sim 3\times 10^8 (s/1km)$ yrs.\footnote{This
    is steeper than the relation of $s^{1/3}$ found by
    \citet{1998AJ....115.2136K}. We suspect that this may relate to
    their initial conditions of super-escape velocity dispersions.}
  The size of the largest bodies that form and the size spectrum are
  comparable in all simulations. More importantly, the efficiency of
  formation ($\epsilon$) remains similarly low in all cases.

\end{itemize}

To summarize, changes in surface density and initial size primarily
affect the conglomeration results through their effect on the mean
collision frequency (eq. \ref{eq:nu}).  However, neither the size
spectrum nor the formation efficiency is much affected by changes in
initial conditions.  

\section{Comparison against Observed Populations}

\begin{figure}[t]
\begin{center}
\includegraphics[width=.45\textwidth, trim = 0 0 0 0, clip, angle=0]{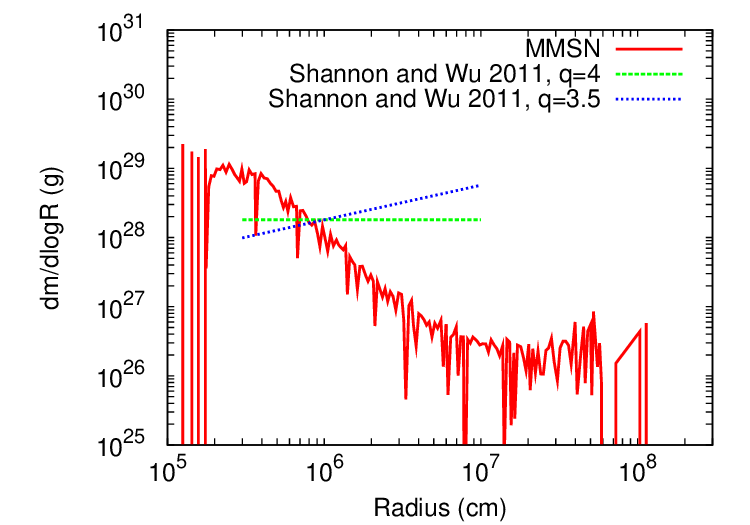}
\caption{ Comparison between the theoretical size distribution
    (this work, MMSN disk) and the inferred size distribution for
    extra-solar debris disks based on their luminosity evolution
    \citet{2011ApJ...739...36S}, the different straight lines are the different size spectra they found to be compatible with the evolution of debris disks. In any case, the theoretical
    calculation produces far too few large bodies, by a few orders
    of magnitude, in the 10-100 km range to account for the bright
    luminosity of these disks at advanced ages (Gyrs).
    The low efficiency of making large bodies by conglomerating
    km-sized sized bodies  is incompatible with the 
    properties of bright extra-solar debris disks.}
\label{fig:plot220}
\end{center}
\end{figure}

 Here, we contrast the conglomeration results against
  observations. There are two contexts where this is possible. First,
  the extra-solar debris disks. The dust brightness of these disks
  reflect the number densities of parent bodies. As the disks age, the
  relevant parent bodies have larger and larger sizes. So by studying
  how the dust luminosities of these disks evolve with time, as an
  ensemble, it is possible to infer the size spectrum as well as
  number densities of the parent bodies
  \citep{2011ApJ...739...36S}. This is illustrated in
  Fig. \ref{fig:plot220}, along with predictions from conglomeration
  calculations. Starting from a MMSN solid disk ($\sim 10 M_\oplus$),
  coagulation of km-bodies leaves most of the primordial mass at the
  initial sizes, and not enough large bodies are formed. This falls
  short to explain the brightness of debris disks at advanced ages, by
  a few orders of magnitudes.

\begin{figure}[h]
\begin{center}
  \includegraphics[width=.45\textwidth, trim = 0 0 0 0, clip, angle=0]{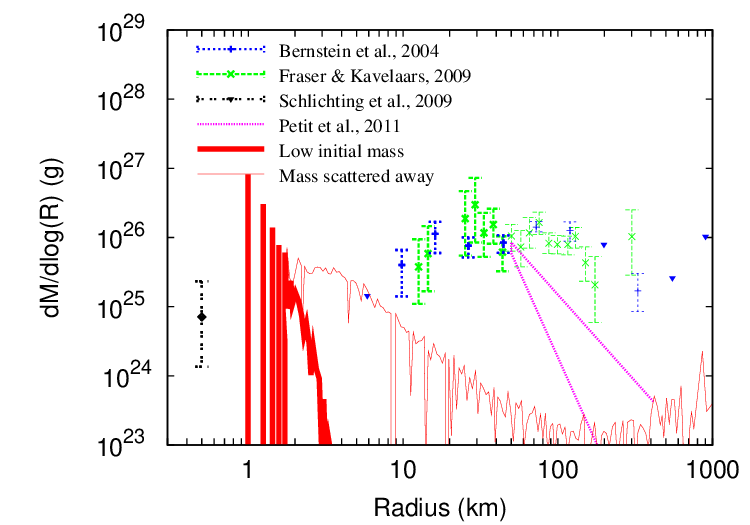}
  \caption{ The growth of Kuiper belt objects after 4.5 Gyrs
      of evolution under two scenario: the thick red line is that
      results from a sparse population of km-sized planetesimals ($0.1
      M_\oplus$) as is constrained by the survival of long period
      Kuiper belt binaries \citep{2012ApJ...744..139P}; while the thin
      red line is that from a MMSN-like solid disk ($10
      M_\oplus$), from which $99.9\%$~of the bodies are then scattered away to make it compatible with the survival of the long period binaries. The observed mass distribution, compiled from
      different surveys, are plotted as points with vertical error
      bars. The survey by \citet{2011AJ....142..131P} specifically
      selected for cold classical objects 
      while those by \citet{Bernstein,2010Icar..210..944F} select
      objects only by their low inclinations and may be contaminated
      by other Kuiper belt populations.  \citet{2009Natur.462..895S} yields an upper limit for Cold Classicals at sub-km size. }
\label{fig:pkcomp}
\end{center}
\end{figure}

 Now we compare our results against observations of Kuiper belt
  objects (Fig. \ref{fig:pkcomp}). We have performed two integrations,
  the first one starting with a low-mass km-sized population ($\sim
  0.1 M_\oplus$), one that is allowed to exist for the age of the
  Solar system without destroying the Kuiper belt binaries
  \citep{2011ApJ...743....1P,2012ApJ...744..139P}. The collision time
  in this case is so long that large bodies cannot be formed, even
  after 4.5 Gyrs of evolution. The second case is our standard MMSN
  disk ($\sim 10 M_\oplus$), which subsequently undergoes a `scattering away' event to make it compatible with the existence of the long period binaries.  Although the un-scattered case is capable of producing the observed
  large Kuiper belt objects, when considering this other constraints, such a scenario is no longer feasible ( also see the introduction).

\section{Comparison with Previous Works}
\label{sec:compare}
\label{sec:comparison}

\subsection{Kenyon \& Luu (1998)}
\label{subsec:ky}

\begin{figure}[h]
\begin{center}
  \includegraphics[width=.45\textwidth, trim = 0 0 0 0, clip, angle=0]{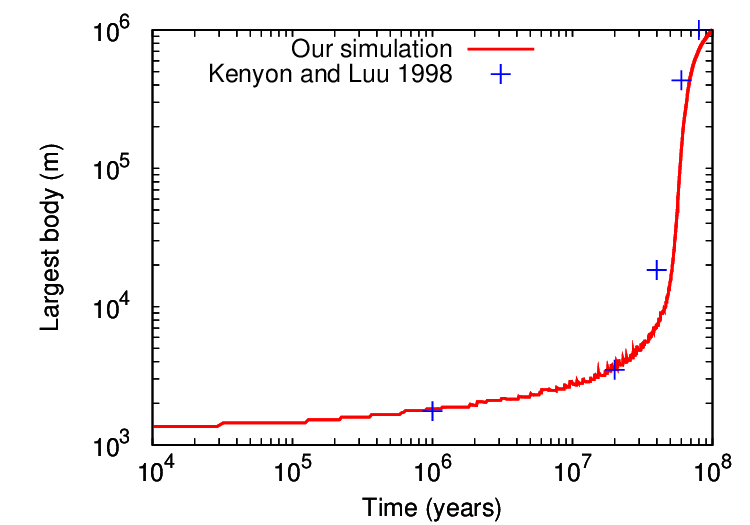}
  \caption{The size of the largest body as a function of time, using
    our conglomeration code (red curve) and from that of
    \citet{1998AJ....115.2136K} (blue crosses).  The growth rates
    agree qualitatively, although ours is slower by $15 \sim 20\%$. }
\label{fig:kl98largestgrowth}
\end{center}
\end{figure}

\begin{figure}[h]
\begin{center}
  \includegraphics[width=.45\textwidth, trim = 0 0 0 0, clip, angle=0]{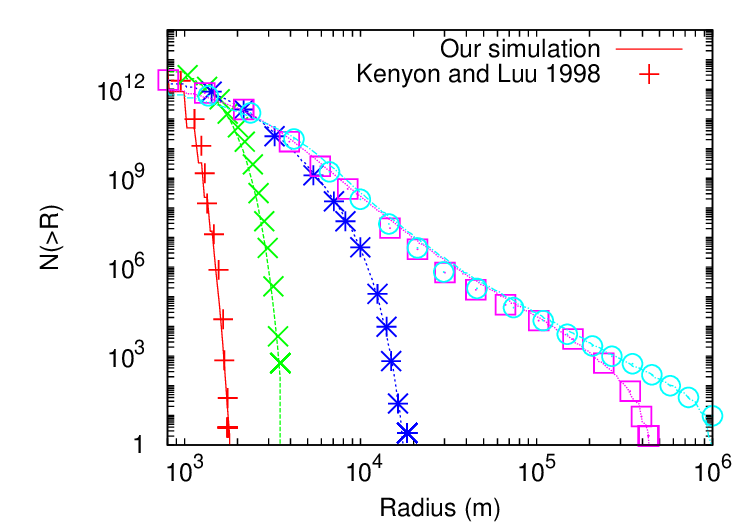}
  \caption{ Comparison of the cumulative size distribution found by
      \citet{1998AJ....115.2136K} (points), to that found in our
      simulation with the same initial conditions (lines). Comparisons
      are plotted when the largest bodies are of the same size, to
      account for the result that our code produces slightly slower
      growth.  This corresponds to figure 9 in that work.}
\label{fig:kl98cnumber}
\end{center}
\end{figure}

We conduct a comparison to the simulations of
\citet{1998AJ....115.2136K}.  We choose this paper for comparison over
subsequent ones from the same group because simulations in this paper
experiences negligible influence from the gas dynamics and have
initial conditions that are the closest to our set-up.  They begin
with bodies of a single size (we will compare here to their $800\m$
case), with a total mass of $10 M_{\oplus}$~in an annulus from $32$ to
$38$ AU and an eccentricity of $e=10^{-3}$.  This value corresponds to
the escape velocity from $10\km$-sized bodies. So the early accretion
(before reaching $10\km$) proceeds slowly without the benefit of
gravitational focussing.  Adopting their initial parameters, our
  conglomeration code yields size of the largest body at any given
time, $R_{\rm max}(t)$, as in Figure \ref{fig:kl98largestgrowth}. The
growth time in our simulations agree within $\sim 20\%$ from those in
\citet{1998AJ....115.2136K}.  We also compare the resulting
  cumulative size-number distribution, and find it to be in good
  agreement (Fig. \ref{fig:kl98cnumber}).  Whatever minor
  differences that exist may be a result of the different approachs
in mass bins, viscous stirring \& dynamical friction, cross sections,
collisional damping, or any combination thereof.  The overall
agreement of the two studies demonstrates that the outcome is
generic and is not significantly affected by the details of the code.

\subsection{\citetalias{SS11}}
\label{subsec:SS11}

While earlier works have performed exhaustive studies of the conglomeration process,
\citetalias{SS11} is the first to present a simple analytical argument
for the numerical results and is thus the most relevant work for
comparison. Here, we critically examine their argument.

\citetalias{SS11} aims to explain the $q=4$ spectrum, obtained by all
works, as a result of big bodies growing by equally accreting 
small bodies and other big bodies. They assume that the
largest bodies grow both by accreting each other and by accreting small
bodies, at equal rates (`equal accretion'). This assumption, which we
examine here, enables them to derive a mass spectrum of $q=4$, with a
(constant) efficiency of large body formation of $\epsilon \sim
\alpha^{3/4}$. These claims appear to be backed up by their simple yet
elegant numerical experiments.

\begin{figure}[h]
\begin{center}
  \includegraphics[width=.45\textwidth, trim = 0 0 0 0, clip, angle=0]{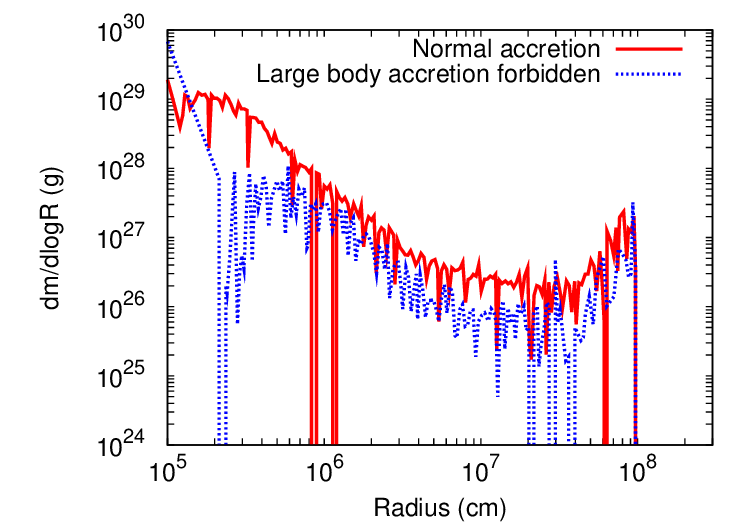}
  \caption{Comparison of the size-number distribution in two
    simulations, plotted when the largest body is $10^3$~km in radius.
    Both simulations begin with $10 m_{\oplus}$ in small bodies,
    distributed with $q = 10$ from $6e15$~to $6e17$~grams in mass.  In
    the solid red line, accretion proceeds as normal.  In the dashed
    blue line, bodies are forbidden from accreting bodies with more
    than $10\%$~of their mass.  Although the latter case cannot
    undergo equal accretion, as postulated by \citetalias{SS11}, it
    still produces a $q \approx 4$~size-number distribution between
    $50$~km and $500$~km.}
\label{fig:funphysics}
\end{center}
\end{figure}

However, it is unclear why, on physical grounds, equal accretion
should occur.\footnote{\citet{2010Icar..210..507O} claimed to observe
  equal accretion in their numerical simulation. However,
  fragmentation included in their study has reduced the mass in, and
  hence accretion from, small bodies.}  Instead, we have shown that
the $q=4$ spectrum arises due to the fact that the largest bodies at
any given time are trans-hill relative to the small bodies. We present
a physical argument why this is naturally achieved (\S
\ref{subsec:three}).  To test whether equal accretion or
  trans-hill accretion is the important physics driving the $q =
  4$~size spectrum, in our code, we disallow big bodies from
accreting one another: they are forbidden from accreting bodies
more than 10\% of their masses. We begin the simulation with a narrow
spectrum (as opposed to a delta-function, as our new formulation will
otherwise stall any growth) of initial bodies.
We recover a $q=4$~size spectrum, despite the absence of
equal accretion (Fig. \ref{fig:funphysics}).

{A second test to judge between the two interpretations is
  possible.  Our expected efficiency of growth scales as $\alpha$,
  while SS11 predicts an efficiency of $\alpha^{3/4}$.  In Figs.
  \ref{fig:alphaone} and \ref{fig:alphathreequarts}, we compare
  simulations conducted at different semimajor axis, and hence
  different $\alpha$, plotted before the onset of oligarchy.  To avoid
  oligarchic effects, collisional cooling, and contributions from
  intermediate-sized bodies, the simulations are started with
  different initial sizes that are proportional to $\alpha$. It
  appears that the numbers of large bodies scale better with $\alpha$
  than with $\alpha^{3/4}$. The numerically obtained efficiency is
  $\sim 3 \alpha$.
This appears to support the interpretation of trans-hill accretion. Interestingly, during trans-hill accretion, accretion of small bodies
and large bodies occurs at a fixed ratio \citep[see Section 5.1
of][]{2014ApJ...780...22L}. But it is an effect rather than a cause.

\begin{figure}[h]
\begin{center}
  \includegraphics[width=.45\textwidth, trim = 0 0 0 0, clip,
  angle=0]{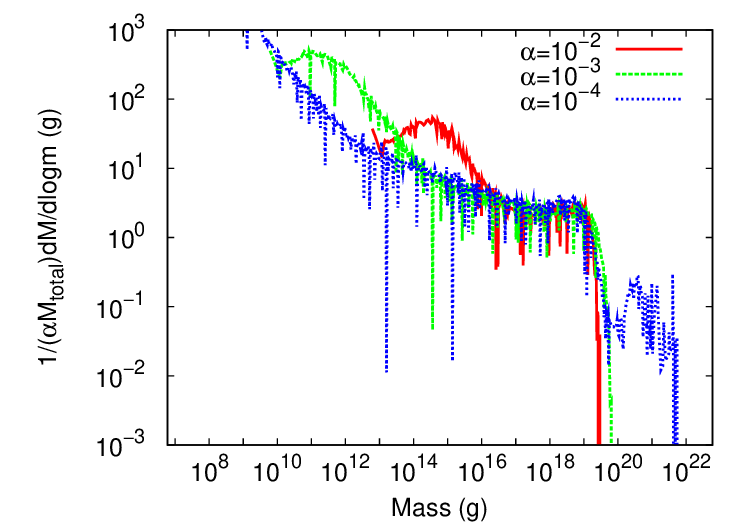}
  \caption{Comparison of the mass distribution for three
      different $\alpha$ values, $\alpha = -2, -3,$~and $-4$.
      Simulations are plotted as slow runaway begins to turn over to
      oligarchic growth.  The $q \approx 4$~spectrum ranges from from
      $\sim 10^{16}$~to $10^{19}$~grams.  The mass per mass decade is
      normalised by $\alpha^{1}$.  This appears to be a better
      scaling than $\alpha^{3/4}$ (Fig. \ref{fig:alphathreequarts}).}
\label{fig:alphaone}
\end{center}
\end{figure}

\begin{figure}[h]
\begin{center}
  \includegraphics[width=.45\textwidth, trim = 0 0 0 0, clip, angle=0]{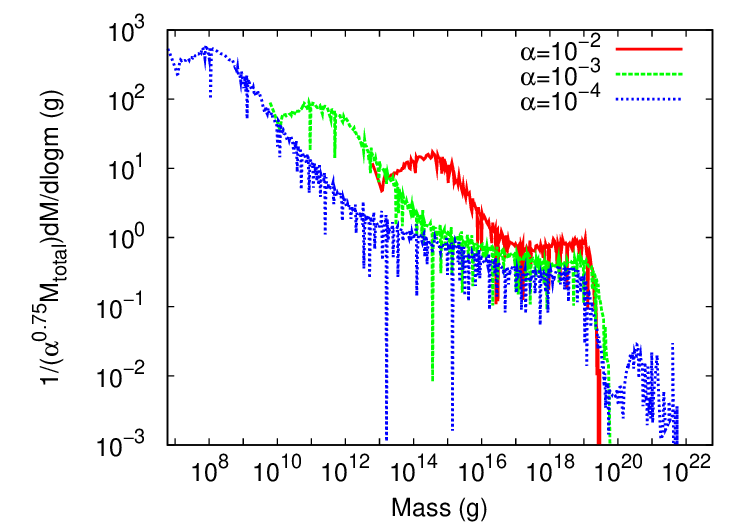}
  \caption{Same as Fig. \ref{fig:alphaone} but with the mass
      spectrum scaled by $\alpha^{3/4}$.}
\label{fig:alphathreequarts}
\end{center}
\end{figure}

\section{Conclusions}

In this contribution, we focus on the efficiency of forming large
bodies by conglomeration, starting from a sea of small planetesimals
that are not cooled by frequent collisions. We find, concuring with
previous studies, that the formation efficiency is $\epsilon \sim
{\mathrm{ a \,\, few}}\times \alpha$, or $\sim 10^{-3}$
at the distance of the Kuiper belt.

We obtain this result numerically by constructing a conglomeration
code that incorporates physical processes such as viscous stirring,
dynamical friction and accretion. This code also has the capability of
dealing with catastrophic disruption and collisional damping. But in
the current study with initial size of $1$ km, collisional cooling is not
relevant and we do not allow bodies to break down to smaller particles.

We also derive the formation efficiency analytically
\citep{2014ApJ...780...22L}.
We find that in collisionless environments, growth passes through
several stages, but naturally transitions to the most critical stage,
which we term 'trans-hill' growth 
Growth is a run-away process when small bodies are super-hill and an
orderly process when small bodies are sub-hill. Consequently, the
biggest body at any given time sits at trans-hill, or $u \sim \vhill
(R_{\rm max})$. 
The fact that the system is driven to trans-hill growth is the key for
why the size spectrum is $dn/dR \propto R^{-q}$ with $q =4$,
and why the formation efficiency  is $\epsilon
\sim {\rm a\,\, few\,}\times \alpha$.

To be confident of our numerical procedure, we have tested individual
components in the code against the order-of-magnitude formulae in
GLS. We have also performed detailed comparisons with previous
simulations. These include \citetalias{SS11} and
\cite{1998AJ....115.2136K}.  All previous works yield a similarly low
formation efficiency, and a similar size spectrum ($q\approx 4$) as we
obtain here. But some detailed differences exist.  Comparison against
\cite{1998AJ....115.2136K} show that we reproduce their overall
  results, but with $\sim 10\%$ differences in both time-scales and
  size distributions.  \citetalias{SS11} postulated that equal
accretion drives the $q=4$~size spectrum, and low efficiency.  We,
  by explicitly forbidding accretion among big bodies, show that the
  correct physical interpretation is likely not related to equal
  accretion, but to trans-hill accretion. This is also supported by
  the scaling of formation efficiency with distance to the central
  star.

In the collisionless limit, trans-hill growth ends when the largest
bodies become oligarchs, each responsible for stirring their own
food. Subsequent accretion in this phase does not increase
significantly the formation efficiency, at least in the Kuiper belt,
due to the impossibly long accretion timescale for Pluto-type objects.

Surface densities in large bodies in the Kuiper belt have been
measured to be $\sim 10^{-3}$ of that of the MMSN, and the largest
bodies are $\sim 1000\km$ in size.  These have traditionally been
viewed as successes for the collisionless coagulation theory. However,
recent discoveries of bright extra-solar debris disks call this into
question. Their dust luminosities reveal that they likely harbor large
bodies that are a factor of $\sim 1000$~times greater in number
than that in our Kuiper belt \citep{2011ApJ...739...36S}, comparable
to the total mass in a MMSN solid disk in these outer regions.  This
conflicts with the low formation efficiency, which is the generic
outcome of collisionless conglomeration.  So even if the size spectrum
of KBOs may be explained by the collisionless process, doubts on which
still linger, the low efficiency of collisionless conglomeration
precludes it from being the process which forms debris disks,
including the Kuiper belt.  In our simulations, km-size bodies reach
such high velocity dispersion towards the end, that their mutual
encounter should cause fragmentation into smaller particles. This is
also observed in \citet{KenyonandLuu:1999}.  So even if planetesimals
start as large as $1\km$, collisional cooling may set in at some
stage. An evolutionary path that is qualitatively different from that
described here may ensue.

To resolve the issue of long formation timescale for Uranus and
Neptune, \citet{Goldreich:2004b} have proposed that conglomeration
proceeds in a collisional environment, where small bodies are so small
they are cooled by frequent collisions. In future works \citep{swl},
we follow this path and demonstrate that collisional conglomeration
would also be able to raise the efficiency of formation to of order
unity, thereby explaining, within one paradigm, the formation of the
Kuiper belt and the extra-solar debris disks.

\acknowledgements

We thank the first referee, Chris Ormel, and a second anonymous
referee for a knowledgeable critique of our numerical procedure. 
  We substantially revamped our numerical procedures following these comments.  YW
acknowledges grants from NSERC and the government of Ontario.  YL acknowledges  grants AST-1109776 and AST-1352369 from NSF, and NNX14AD21G from NASA.  AS was supported by the
government of Ontario by a Ontario Graduate Scholarship in Science and
Technology; and is supported by the European Union through ERC grant
number 279973.

\bibliographystyle{apj}
\bibliography{coagreborn}

\end{document}